
\magnification=1100
\tolerance=100000
\hyphenpenalty=1000
\raggedbottom

\def\ainit{\hoffset=.0 truecm
            \voffset=1. truecm
            \hsize=17. truecm
            \vsize=23.5 truecm
            \baselineskip=11.pt
            \lineskip=0pt
            \lineskiplimit=0pt}

\ainit
%
%

\def\Msol{\,{\rm M_\odot}}
\def\Lsol{\,{\rm L_\odot}}

\def\Msyr{\,{{\rm M_\odot}\,{\rm yr}^{-1}}}
%
%

\def\lsim{\,\lower2truept\hbox{${< \atop\hbox{\raise4truept\hbox{$\sim$}}}$}\,}
\def\gsim{\,\lower2truept\hbox{${> \atop\hbox{\raise4truept\hbox{$\sim$}}}$}\,}

%
%
\def\oneskip{\vskip\baselineskip}
\centerline{\null}
\oneskip
\noindent

\centerline{ DUST IN HIGH REDSHIFT RADIO GALAXIES AND}

\centerline{ THE EARLY EVOLUTION OF SPHEROIDAL GALAXIES}
\bigskip
\centerline{\bf Paola Mazzei and Gianfranco De~Zotti}

\medskip
\centerline{Osservatorio Astronomico, Vicolo dell'Osservatorio 5,
I-35122 Padova, Italy}

\centerline{e-mail: mazzei@astrpd.pd.astro.it}

\bigskip
\bigskip
\noindent
{\bf ABSTRACT}

\medskip\noindent
Several lines of evidence suggest that early-type galaxies might have been
very dusty during their initial evolutionary stages, characterized by
intense star formation activity. The radio selection has provided, by far,
the richest samples of high-$z$ galaxies, most likely of early type, which
may therefore yield crucial direct information on the period when the
bulk of stars were born. In order to investigate the role of dust in
these phases, we have analyzed recent observations of a number of
high-$z$ radio galaxies in the framework of a self-consistent
scenario for the evolution of early type galaxies.
The available data can be fully accounted for by ``opaque'' models similar to
that already used by Mazzei and De Zotti (1994a) to fit the spectrum of the
hyperluminous galaxy IRAS~F$10214+4724$. However, there is still considerable
latitude for models. Good fits can be obtained with galactic ages
ranging from 0.05 to 2 Gyr. Crucial constraints may be provided by
ground-based sub-mm measurements and by observations with the forthcoming
Infrared Space Observatory (ISO).

\bigskip\noindent
{\bf Key words:} galaxies: evolution -- galaxies: elliptical and lenticular --
infrared: galaxies.

\bigskip

\noindent
{\bf 1 INTRODUCTION}

\medskip\noindent
The evolution with cosmic time of stellar populations in galaxies can be
directly investigated by looking at the spectral energy distribution (SED)
of galaxies at different redshifts. ``Normal'' high-$z$ galaxies,
however, proved to be remarkably elusive: none of the 100 (out of a total of
104) galaxies with measured redshift in the sample of Colless et al. (1993),
complete to $B=22.5$, has $z>0.7$; the highest redshift in the complete
$B <24$ sample of Cowie et al. (1991) is $z=0.73$; the 73\% complete redshift
survey to $B=24$ by Glazebrook et al. (1995) has discovered only one
galaxy at $z>1$ ($z=1.108$); the
measured redshifts of near-IR selected galaxies down to $K=20$ (nearly 100\%
complete at $K<18$ and $\sim 70\%$ complete at $K=19$--20) are $\lsim 1$,
with the exception of one object at $z=2.35$.

So far, the only effective method for finding high-$z$ galaxies has been
optical identifications of radio sources (Chambers, Miley \& van Bruegel 1988;
Lilly 1988; Chambers, Miley \& van Bruegel 1990; Rawlings et al. 1990;
Windhorst et al. 1991; McCarthy 1993; Lacy et al. 1994; Dey, Spinrad \&
Dickinson 1995; van Ojik et al. 1994).
Of course, the interpretation of the observed SED is complicated by the need
of disentangling the starlight from the emission of the active nucleus,
and of properly allowing for effects induced by nuclear activity, e.g.
star formation triggered by the radio jet as it propagates outwards from the
nucleus, which may be responsible for the observed close alignment of optical,
infrared and radio emissions (see McCarthy 1993, for a recent discussion).

Nevertheless, high-$z$ radio galaxies can provide unique information on early
phases of galaxy evolution, when intense star formation occurred. As pointed
out by several authors, the failure of optical searches to detect primeval
early-type galaxies (Djorgovski, Thompson \& Smith 1993; De Propris et al.
1993; Parkes, Collins \& Joseph 1994) might indicate that they were strongly
obscured by dust.  Franceschini et al. (1994) have shown that under the
assumption that ellipticals were very dusty during the phases of most intense
star formation, a consistent picture can be obtained, accounting for the
scarcity of high-$z$ optically selected galaxies as well as for the deep counts
of galaxies at far-IR and radio wavelengths.

The presence of large amounts of dust in high redshift radio galaxies has
indeed been indicated by a number of recent observations (Eales \& Rawlings
1993; Dunlop et al. 1994; van Ojik et al. 1994; Dey et al. 1995; Carilli
1995). The interpretation
of the observed SEDs, however, is not straightforward since the
commonly used population synthesis models do not allow for dust extinction.

In this paper we exploit the evolutionary population synthesis models
for early type galaxies worked out by Mazzei, De Zotti \& Xu (1994), taking
into account in a self consistent way dust absorption and re-emission,
to analyze the continuum spectra of high-$z$ radio galaxies for which enough
data are available; a non-thermal contribution is also included.
In \S~2 we briefly describe the models; in \S~3 we present
the fits to observed continuum spectra; in \S~4 we discuss and summarize the
main results.

Throughout this paper we adopt $H_0=50\,\hbox{km}\,\hbox{s}^{-1}\,
\hbox{Mpc}^{-1}$.

\bigskip
\noindent
{\bf 2. MODELLING THE SED FROM UV TO FAR--IR}

\medskip\noindent
Following Mazzei et al. (1994), the evolution of the synthetic SED has been
linked (albeit roughly) to chemical evolution, so that the increased
metallicity of successive stellar generations is taken into account.
The galaxy is assumed to be a close system, i.e. both galactic winds and
inflow of intergalactic gas are neglected. Matteucci (1992) found that,
in the presence of a dynamically dominant dark matter component
distributed like
stars, elliptical galaxies develop winds very late; and, in general,
we expect winds to be important particularly for low mass galaxies, while
we will be dealing here with large objects.

We have adopted Schmidt's (1959) parametrization of the
star--formation rate (SFR), $\psi (t)$:
$$\psi (t) = \psi _0 \left(m_{b}\over 10^{11}\Msol\right)
f_g^n\ \Msyr , \eqno(1)$$
where $f_g = m_{\rm gas}/m_{b}$ is the fractional mass of gas in
the galaxy ($m_b$ is the total mass in baryons),
assumed to be, initially, unity. Current analyses favour $n=1$--2 (Buat 1992),
but $n=0.5$ has also been advocated (Madore 1977). Successful models for
early type galaxies require $\psi_0 \geq 100\,\Msol\,\hbox{yr}^{-1}$
for $n \geq 1$ or $\psi_0 \geq 35\,\Msol\,\hbox{yr}^{-1}$
for $n = 0.5$.

As for the initial mass function (IMF), $\phi (m)$,
we have adopted a Salpeter (1955) form:
$$\phi (m) dm = A \left({m\over \Msol}\right)^{-2.35} d\left({m\over
\Msol}\right)\qquad m_l \leq m \leq m_u, \eqno(2)$$
with $m_u=100\,\Msol$ and $m_l$ ranging from $0.01\,\Msol$ to $0.5\,\Msol$.

Stellar lifetimes are taken into account (i.e. we have not resorted to the
instantaneous recycling approximation). The gas is assumed to be well mixed
and uniformly distributed. The variations with galactic age, $T$, of
the fractional gas mass $f_g(T)$ and of the gas metallicity $Z_g(T)$ are
obtained by numerically solving the standard equation for chemical
evolution (for details, see Mazzei, Xu \& De Zotti 1992).

The synthetic spectrum of stellar populations as a function of the galactic
age has been obtained using the theoretical isochrones given by
Bertelli et al. (1990), as extended by Mazzei (1988) up to 100 $\Msol$ and
to an age of $10^6$ yr. Full details can be found in Mazzei et al.
(1992, 1994).

The effect of dust has been taken into account based on the following
assumptions: the dust to gas ratio is proportional to the gas metallicity
$Z_g(T)$; the ``standard'' grain model (Mathis, Rumpl \& Nordsiek 1977; Draine
\& Lee 1984), incorporating a power law grain size distribution, holds at any
time; the size distribution of PAH molecules obeys the power law proposed by
Puget et al. (1985), while their absorption cross sections and emission
properties are those given by Puget \& L\'eger (1989).

Corrections for internal extinction have been computed assuming the
elliptical galaxies to be spherically symmetric, with a density profile
described by King's (1966) formula, and adopting the interstellar
extinction curve given by Seaton (1979) for $\lambda \leq 0.37\,\mu$m
and by Rieke \& Lebofsky (1985) at longer wavelengths. The optical
depth has been normalized requiring that the far-IR to blue luminosity
ratio at the present time be equal to the mean value, $L_{\rm FIR}/L_{\rm B}
\simeq 7\times 10^{-3}$, derived by Mazzei \& De Zotti (1994b) for a
complete sample of local elliptical galaxies. Its
evolution follows directly from that of the gas fraction and of the
metallicity.

The diffuse
dust emission spectrum is modelled taking into account two components: warm
dust, located in regions of high radiation intensity (e.g., in the neighborhood
of OB clusters) and cold dust, heated by the general interstellar radiation
field. Their relative normalization at the present time is obtained fitting the
resulting spectrum to the mean local far-IR colours of ellipticals (Mazzei \&
De Zotti 1994b).  The warm to cold dust emission ratio is
assumed to be proportional to the star formation rate; warm dust is therefore
dominant in the early evolutionary phases considered here.
 PAH molecules are mixed with both these components, as discussed in Mazzei et
al. (1992).

The model also allows for emission from hot circumstellar dust, emitting at
mid-IR wavelengths. Based on
the analysis by Ghosh et al. (1986), we assume that the dominant contribution
comes from stars in the final stage of evolution along the asymptotic giant
branch.

Different
choices for the initial SFR, $\psi_0$, and for the index $n$ [eq. (1)],
all consistent with the observed properties of early type
galaxies, result in very different histories for
the effective optical depth, $\tau_{\rm eff}$, defined by
$L_{\rm FIR}/L_{\rm bol} = 1 -\exp(-\tau_{\rm eff})$, $L_{\rm bol}$ being
the bolometric luminosity of the galaxy. This is because $\tau_{\rm eff}$
is proportional to the product of $f_g(T)$ and $Z_g(T)$ (as a consequence
of the assumption of a dust to gas ratio proportional to the gas metallicity),
the first of which is rapidly decreasing during the early evolution of the
galaxy, while the other is rapidly increasing.

Therefore, relatively small differences in the evolution of the synthetic
spectrum of stellar populations, may translate into drastically different
amounts of dust absorption and re-emission. As illustrated by Fig.~1,
models fitting the observed local properties of early type galaxies and
yielding, at the present time, an effective optical depth equal to the
mean value found by Mazzei \& De Zotti (1994b), may either be optically
thin throughout the entire evolutionary history of the galaxy (although
dust absorption is in any case important during early phases), or
become optically thick for a short time, or even experience an extended,
extremely opaque phase, when essentially all light comes out at far-IR
wavelengths. Models of the latter type provide a very good fit to the
observed SED of the hyperluminous galaxy IRAS~F$10214+4724$ (Mazzei \&
De Zotti 1994a), although the recent observational evidences that this source
is gravitationally lensed (Matthews et al. 1994; Broadhurst \& Lehar 1995;
Graham \& Liu 1995; Trentham 1995; Eisenhardt et al. 1995;
Serjeant et al. 1995; Downes et al. 1995) complicate the interpretation
of the spectral energy distribution. In fact, differential magnification
of different components having different sizes may change the source colours.
As pointed out by Downes et al. (1995), it is likely that the near- and
mid-IR emissions are more amplified than the far-IR emission.

In the following discussion we will refer only to models with $n=0.5$
and $n=1$. We stress that, in the present data situation, models
are necessarily rough and are to be taken only as
descriptive of a possible astrophysical situation. For example, we regard
$n=0.5$ models as a simple way to describe a situation whereby galaxies become
strongly obscured during some evolutionary phase; the physical reason for that,
however, might be related not to this particular form of the SFR, but, e.g., to
merging of galaxies with a metal enriched interstellar medium.

The models successfully match the observed SEDs of local galaxies over four
decades in frequency (Mazzei et al. 1992, 1994; Mazzei, Curir \& Bonoli 1995).
As illustrated by Fig. 2a, where predictions corresponding to
initial SFR's, $\psi_0$, decreasing from values typical of ellipticals (top)
to values appropriate to irregulars (bottom) are shown,
they are also nicely consistent with the observed colour distribution
of galaxies in the complete $K < 20$ sample of Songaila et al. (1994); the
reddest observed colours are close to predictions of dusty models for early
type galaxies (solid lines).

Incidentally we note that, according to our models, evolutionary effects cannot
be neglected for $z \geq 1$; in fact, the colours predicted by models are
bluer than those expected by
simply redshifting the SEDs of template
local galaxies, due to the higher SFR. Therefore photometric redshift estimates
hinging upon the SEDs of unevolved galaxies may be misleading.

The predicted colour distribution (Fig. 2) is rather sensitive to the adopted
cosmological model. For given formation redshift $z_f$, low density models
imply considerably redder colours (cf. Fig 2c,f); for example, for $z_f=10$
($t_f = 0.4\,$Gyr if $q_0=0.5$ or $1.8\,$Gyr if $q_0=0$),  we
expect values of  $(B-K)$ and $(I-K)$ of
up to $\simeq 9.6$ and 7.7 respectively
if $q_0=0$, while $(B-K)\leq 8.8$ and $(I-K)\leq 5.5$ if $q_0 = 0.5$.
We note, however, that the colour distribution is also sensitive to $z_f$;
lower values of $z_f$ entail bluer $(B-K)$ and $(I-K)$ (Fig. 2b,e).

As illustrated by Fig. 3, our models satisfactorily account for the increasing
spread of K-band magnitudes of radiogalaxies at $z \geq 1$ in terms of the
combined effect of evolution of stellar populations and of dust extinction.

All in all, rather general arguments lead to the conclusion that dust
extinction plays an important role during the early evolution of
elliptical galaxies; there are also indications that a large fraction
of ellipticals may go through an opaque phase (Franceschini et al. 1994).
As pointed out by several authors, the inferred ages of high-$z$
radiogalaxies might be reduced if there is significant reddening (cf.
McCarthy 1993), thus alleviating the problem of galaxy ages uncomfortably
high in comparison with the age of the universe as well as the difficulty
to understand the alignment between optical/infrared continua and the
(presumably short lived) radio jets. The possibility of
reproducing the observed SED of high-$z$ radio galaxies in terms of
a young reddened starburst has been questioned, however (cf. Chambers et al.
1990).

In the following section we will analyze the observations of the most
extensively studied high redshift radiogalaxies to show that, in fact,
they are fully consistent with the presence of strong dust extinction.

\bigskip

\noindent
{\bf 3. FITTING THE OBSERVED SEDs OF HIGH-$z$ RADIO GALAXIES}

\medskip\noindent
In interpreting the photometric data on radio galaxies it is necessary to take
into account that the observed luminosity is a mix of stellar and nuclear
emissions. As for the shape of the continuum spectrum of the latter, the
mean SED of local AGNs, derived by Granato \& Danese (1994), has been adopted.
It is also necessary to take into account that low resolution continuum
spectra may be distorted by contributions of intense emission lines in the
broad-band filters (Eisenhardt \& Dickinson 1992; Eales et al. 1993b;
Eales \& Rawlings 1993).

We discuss here the four best observed $z>2$ radiogalaxies, namely:
4C41.17, the second highest redshift galaxy known, B2~$0902+343$,
the first $z>3$ galaxy discovered, 6C$1232+39$, and $53W002$.

In general, after subtraction of the emission line contributions to broad-band
fluxes, the SED turns out to be consistent with a young starburst,
of age in the range 0.05--0.3 Gyr, assuming that the UV continuum is
dominated by stellar emission. If, on the other hand,
the AGN provides most of the rest-frame emission at 1000{\AA}, stellar
populations must be older and galactic ages as large as 1--$2\,$Gyr
are derived.

In any case, there is considerable latitude for models, implying, among
other things, that little can be inferred from optical/near-IR photometry alone
on the amount of dust extinction. Millimeter and sub-mm observations of
4C41.17, B2~$0902+343$, and $53W002$ imply large masses of dust
and the upper limits on 6C$1232+39$
are still compatible with a similar situation. If so, the
total baryonic masses and the bolometric luminosities must be large:
$m_b > 10^{12}\Msol$, $L_{\rm bol} > 10^{13}\Lsol$.

\medskip\noindent
{\it 3.1. 4C41.17}
\medskip\noindent
The published data on this $z=3.8$ radio galaxy, the second most distant
galaxy known, are shown in Fig. 4. The $2.2\,\mu$m flux is from the
line-free $K_S$ magnitude in a 4'' diameter beam measured by Graham et al.
(1994). No correction for contamination by emission lines is possible for
I-band and R-band fluxes measured by Chambers et al. (1990); spectroscopic
data of the same authors, however, indicate that the observed magnitudes
can safely be taken as representative of the continuum emission. The two
points at the shortest wavelengths are from a low dispersion spectrum
by Chambers et al. (1990).

This radiogalaxy has been clearly detected at mm and sub-mm wavelengths
(Dunlop et al. 1994; Chini \& Kr\"ugel 1994). As discussed in both
these papers, it is highly implausible that the observed emission is
of non-thermal origin. The present models straightforwardly account for
it in terms of dust re-radiation,
dominated by warm dust, heated by the hot, young stars. The
available data, however, do not strongly constrain the warm dust temperature,
$T_w$. Shown in Fig.~4 are predictions assuming $T_w = 60\,$K (as in the case
of IRAS~F$10214+4724$, cf. Mazzei \& De Zotti 1994a) and $T_d = 46\,$K.

Although sub-mm observations were made using a
beam size of 16.5'' (Dunlop et al. 1994) or of 11'' (Chini \& Kr\"ugel 1994),
much larger than the apertures for optical/IR photometry,
no aperture corrections have been applied to any data. From an image of
the region of Ly$\alpha$ obscuration, Dunlop et al. (1994)
conclude that the dust is concentrated in a region of a few arcsec in diameter
centered on the flat-spectrum core of the radiogalaxy; any aperture effects
should therefore be small compared with the other uncertainties.

The $3\sigma$ IRAS upper limits at 12 and $25\,\mu$m derived by Dunlop et al.
(1994) are at or above the upper border of the figure.

Panel a) of Fig.~4 illustrates fits by ``young'' models. The galactic ages
range from 0.05 to 0.1 Gyr; thus, 4C41.17 may really be a protogalaxy, in the
sense that only a small fraction of stars have yet formed; in fact, our models
yield a gas fraction $f_g \simeq 0.9$. The best fit non-thermal contribution
(solid line) is $\simeq 35\%$--$50\%$ of the continuum emission at $\lambda =
1000\,${\AA}.

The gas metallicity, $Z_g$, and the total baryonic mass, $m_b$, also depend on
the lower mass limit, $m_l$, of the Salpeter IMF. For $0.01 \leq m_l/\Msol \leq
0.5$ we have $.0015 \leq Z_g \leq 0.01$, the lowest values corresponding to the
lowest $m_l$ and to the youngest ages, and  $1.85 \leq m_b/10^{12}\Msol \leq
6.5$;  of course, much lower baryon masses  be obtained adopting a
substantially
higher $m_l$.
The bolometric luminosity is in the range 1.2--$3.6\times 10^{13}\Lsol$. Most
of the uncertainty follows from the lack of far-IR data.
Measurements with the forthcoming
Infrared Space Observatory (ISO) should be able both to assess the bolometric
luminosity and to strongly constrain the dust temperature distribution.

The internal reddening, in the observer's frame, implied by the models
is $E(R-K)=1.0 $--1.6 (the largest value corresponding to
the highest value of $T$),
consistent with Eales and Rawlings's (1993) estimate, $E(R-K)=1.6$, based on
line ratios. Models imply a very high SFR, ranging from
about 5000 up to $\simeq\, 8000\Msyr$, where the lowest value
correspond to the highest $m_l$.

Figure 4b illustrates the predicted SED in the case of a galactic age of 2
Gyr. In this case, the AGN must dominate the 1000{\AA} rest-frame emission.
The inferred baryonic masses ($4 \leq m_b/10^{12}\Msol \leq 8.6$) are larger
than for ``young'' models. The residual gas fraction is small ($f_g
\simeq 0.0015$--0.06) and the predicted colour excess, $E(R-K)$, ranges from
0.44 to 0.9 mag with a SFR ranging from about 400 to
$2100 \Msyr$.

\medskip\noindent
{\it 3.2. B2~$0902+343$}

\medskip\noindent
Shown in Fig.~5 are the optical observations of Lilly (1988),
the K-band flux, after subtraction of the line emission, estimated by
Eales et al. (1993b; see also Eisenhardt \& Dickinson 1992),
the $800\,\mu$m upper limit by Hughes, Dunlop \& Rawlings (1995)
and the detection at 1.3 mm by Chini \& Kr\"ugel (1994).

Evidences of strong extinction have been reported by Eales \& Rawlings (1993),
who estimated $E(R-K)=2.1\,$mag from line ratios, and by
Eisenhardt \& Dickinson (1992), who pointed out that the large
[OIII]/Ly$\alpha$ ratio, the discrepancy between the positions of the
radio core and the UV/optical emission peaks, and the strong spatial
variations in the $R-K$ colour could all be interpreted as due to
internal extinction.

Panels a) and b) of Fig. 5 show fits by ``young'' and somewhat older models,
respectively. Fits were found only for ages of less than 1 Gyr.

\medskip\noindent
{\it 3.3. 6C$1232+39$}

\medskip\noindent
Deep optical, infrared, and radio images of this $z=3.22$ radio-galaxy
have been
obtained by Eales et al. (1993a). The K-band continuum flux is very uncertain:
by comparing their spectroscopy with published photometry,
Eales \& Rawlings (1993) derived an estimate of  $34\pm 10\,\mu$Jy,
whereas
there is no evidence of a continuum flux in their infrared spectrum.
Plotted in Fig.~6 is the $3\sigma$ upper limit ($37\,\mu$Jy) on
the continuum emission measured by Eales \& Rawlings (1993) from the
spectrum.
Eales et al. (1993a) point out that the intensity peaks of B and R images
are offset from the K-band peak, which coincides with the center of
the radio image; they suggest that this difference in structure may be due to
dust obscuring the emission from the center of the galaxy at the shorter
wavelengths. The upper limits at $800\,\mu$m and at $1.3\,$mm are
from Hughes et al. (1995) and Chini \& Kr\"ugel (1994), respectively.

No estimates of the non-thermal contribution to the observed fluxes are
available. Again, ``young'' models, suggesting $0.05\le T(\hbox{Gyr}) \le 0.2$,
imply a minor ($\sim 10\%$)
contribution from the AGN at $\lambda_{\rm rest frame} = 1000\,${\AA};
if this contribution is increased to, say, $\simeq 60\%$, fits require a
somewhat older age, but still $< 1\,$Gyr, as in the case of
B2~$0902+343$.

Because of
the lack of a reliable estimate of both the dust extinction and the far-IR
emission, models allow the possibility of either small or large
(up to $E(R-K) = 1.7\,$mag) dust extinction.

\medskip\noindent
{\it 3.4. 53W002}

\medskip\noindent
This $z=2.39$ radio-galaxy (Fig. 7) is the most distant so far discovered
by the LBDS survey (Windhorst et al. 1991). HST WFC V-band and I-band
images (Windhorst et al. 1992; 1994) show a radial
intensity profile consistent with that of an elliptical galaxy, plus an
unresolved central source (AGN) containing $30 \pm 10$\% of the total light.
Under the plausible assumption that the galaxy surface brightness is
monotonic with radius, Windhorst et al. (1992) conclude that the AGN
contributes at most 20\% in V ($\lambda_{\rm rest} \simeq 1600\,${\AA})
and 25\% in I ($\lambda_{\rm rest} \simeq 2600\,${\AA}), with a secure upper
limit of 37\%. The ground based spectroscopy of Windhorst et al. (1991)
indicates a non thermal fraction at $\lambda_{\rm rest} = 1450\,${\AA} of
$35\pm 15$\%.

The near-IR spectrum obtained by Eales \& Rawlings (1993) is very noisy
and the correction for the line contribution in the K-band is correspondingly
very uncertain. We have adopted the line-corrected magnitude given in Table
2 of Windhorst et al. (1994).

Assuming that the observed low value of the ratio Ly$\alpha/$H$\alpha$ is
due to dust extinction, Eales \& Rawlings (1993) obtained a quite
substantial dust reddening, $E(R-K)\simeq 2.4$. The presence of a large
amount of dust may be supported by the (marginal) detection of this galaxy
at $800\,\mu$m by Hughes et al. (1994).
Our models are consistent with $0.75 \leq E(R-K) \leq 1.7$.
Good fits to the observed SEDs are obtained for ages down to 0.1 Gyr. However,
its $a^{1/4}$-like light profile observed by Windhorst et al. (1994)
indicates dynamical relaxation and, therefore, this galaxy
cannot be too young. In Fig. 7 we show models with ages of 0.6 and 1 Gyr.
The assumed AGN contribution ($\simeq 40\%$ in V) is consistent with the
results of ground based spectroscopy by Windhorst et al. (1991) and close to
the upper limit derived from HST images by Windhorst et al. (1992). Note,
however, that the quality of the fit is not crucially dependent on the
AGN emission.
\bigskip
\noindent
{\bf 4. CONCLUSIONS}

\medskip\noindent
Our analysis of the observed SEDs of the best studied $z>2$ radiogalaxies
shows that all the ingredients (stellar population synthesis,
dust extinction, contribution of the AGN) need to be taken simultaneously
into account. Ignoring one of them may lead one astray.

In general, the available data are consistent with galactic
ages ranging from $\sim 0.1$ to 2 Gyr. Young models require small AGN
contributions in the UV; if, on the other hand, stellar populations are
1--2 Gyr old, the UV light may be mostly produced by the AGN.

Young models are consistent with (although do not require) large amounts
of dust extinction such as that indicated by the $800\,\mu$m flux detected
from 4C41.17 (Dunlop et al. 1994), by mm measurements
by Chini \& Kr\"ugel (1994), and by the possible detection at $800\,\mu$m
of 53W002 by (Hughes et al. 1995). The presence of
substantial amounts of dust is also suggested by the low values
of  Ly$\alpha/$H$\alpha$ and Ly$\alpha/$H$\beta$ ratios measured for
several high-$z$ radio-galaxies (Eales \& Rawlings 1993).

The existence of a dusty phase in the evolution of early type galaxies
has been advocated to explain the remarkable scarcity of high-$z$ galaxies
in faint optically selected samples as well as the strong
cosmological evolution in the far-IR suggested by deep IRAS counts
at $60\,\mu$m (Franceschini et al. 1994).

High sensitivity sub-mm and mm observations such as those carried out by
Dunlop et al. (1994), Hughes et al. (1995) and Chini \& Kr\"ugel (1994),
as well as measurements
with the long wavelength photometer (PHOT) on the
forthcoming ISO satellite might detect the corresponding dust re-emission
or, at least, set significant constraints on the models.

\vfill\eject

%
\def\aa #1 #2{{A\&A,}~{#1}, {#2}}
\def\aas #1 #2{{A\&AS,}~{#1}, {#2}}
\def\araa #1 #2{{ ARA\&A,}~{ #1}, {#2}}
\def\aj #1 #2{{ AJ,}~{ #1}, {#2}}
\def\alett #1 #2{{ Ap. Lett. \& Comm.,}~{ #1}, {#2}}
\def\apj #1 #2{{ ApJ,}~{ #1}, {#2}}
\def\apjs #1 #2{{ ApJS,}~{ #1}, {#2}}
\def\ass #1 #2{{ Ap\&SS,}~{ #1}, {#2}}
\def\baas #1 #2{{ BAAS,}~{ #1}, {#2}}
\def\ca #1 #2{{ Comm. Ap. Sp. Phys.}~{ #1}, #2}
\def\fcp #1 #2{{ Fundam. Cosmic Phys.}~{ #1}, {#2}}
\def\memsait #1 #2{{ Memorie Soc. Astr. Ital.}~{ #1}, {#2}}
\def\mnras #1 #2{{ MNRAS,}~{ #1}, {#2}}
\def\qjras #1 #2{{ QJRAS,}~{ #1}, {#2}}
\def\nat #1 #2{{ Nat,}~{ #1}, {#2}}
\def\pasj #1 #2{{ PASJ,}~{ #1}, {#2}}
\def\pasp #1 #2{{PASP,}~{ #1}, {#2}}

\def\sovastr #1 #2{{ SvA,}~{ #1}, {#2}}
\def\asr #1 #2{{ Adv. Space Res.}~{ #1}, {#2}.}
\def\ssr #1 #2{{ Space Sci. Rev.}~{ #1}, {#2}.}
\def\va #1 #2{{ Vistas in Astronomy}~{ #1}, {#2}}

\def\book #1 {{ ``{#1}'',\ }}

\def\ref{\noindent\hangindent=20pt\hangafter=1}

\noindent
{\bf REFERENCES}

\bigskip

\ref
Bertelli P., Betto R., Bressan A., Chiosi C., Nasi E., Vallenari A., 1990,
\aas 85 845

\ref
Broadhurst T., Lehar J., 1995, ApJ, in press

\ref
Buat V., 1992, \aa 264 444

\ref
Carilli  C.L., 1994,  A\&A submitted

\ref
Chambers K.C., Miley G.K., van Breugel W., 1988, \apj 327 L47

\ref
Chambers K.C., Miley G.K., van Breugel W., 1990, \apj 363 21

\ref
Chini R., Kr\"ugel E., 1994, \aa 288 L33

\ref
Colless M., Ellis R., Taylor K., Hook R., 1993, \mnras 261 19

\ref
Cowie L.L., Songaila A., Hu E.M., 1991, \nat 354 460

\ref
De Propris R., Pritchet C.J., Hartwick D.A., Hickson P., 1993, \aj 105 1243

\ref
Dey A., Spinrad H., Dickinson M., 1995, \apj 440 515

\ref
Djorgovski S., Thompson D., Smith J.D., 1993, eds. C. Akerlof, M. Srednicki,
Texas/PASCOS '92: Relativistic Astrophysics and Particle Cosmology,
Ann. N.Y. Acad. Sci. 688, 515

\ref
Downes D., Solomon P.M., Radford S.J.E., 1995, preprint

\ref
Draine B.T., Lee H.M., 1984, \apj 285 89

\ref
Dunlop J.S., Hughes D.H., Rawlings S., Eales S.A., Ward M.J., 1994, \nat
370 347

\ref
Eales S.A., Rawlings S., 1993, \apj 411 67

\ref
Eales S.A., Rawlings S., Dickinson M., Spinrad H., Hill G.J., Lacy M., 1993a,
\apj 409 578

\ref
Eales S.A., Rawlings S., Puxley P., Rocca-Volmerange B., Kuntz K., 1993b,
\nat 363 140

\ref
Eisenhardt P., Dickinson M., 1992, \apj 399 L47

\ref
Eisenhardt P., Soifer B.T., Armus L., Hogg D., Neugebauer G., Werner M.,
1995, ApJ, in press

\ref
Franceschini A., Mazzei P., De Zotti G., Danese L., 1994, \apj 427 140

\ref
Glazebrook K., Ellis R., Colless M., Broadhurst T., Allington-Smith J.,
Tanvir N., 1995, \mnras 273 157

\ref
Ghosh S.K., Drapatz S., Peppel U.C., 1986, \aa 167 341

\ref
Graham J.R., Liu M.C. 1995, \apj 449 L29

\ref
Graham J.R., Matthews K., Soifer B.T., Nelson J.E., Harrison W., Jernigan
J.G., Lin S., Neugebauer G., Smith G., Ziomkowski C., 1994, \apj 420 L5

\ref
Granato G.L., Danese L., 1994, \mnras 268 235


\ref
Hughes D.H., Dunlop J., Rawlings S., 1995, preprint

\ref
King I., 1966, \aj 71 64

\ref
Lacy M., Miley G., Rawlings S., Saunders R., Dickinson M., Garrington S.,
Maddox S., Pooley G., Steidel C.C., Bremer M.N., Cotter G., van Ojik R.,
R\"ottgering H., Warner P., 1994, \mnras 271 504

\ref
Lilly S.J., 1988, \apj 333 161

\ref
Madore B.F., 1977, \mnras 178 1

\ref
Mathis J.S., Rumpl W., Nordsieck K.H., 1977, \apj 217 425

\ref
Matteucci F., 1992, \apj 397 32

\ref
Matthews K., Soifer B.T., Nelson J., Boesgaard H., Graham J.R., Harrison W.,
Irace W., Jernigan G., Larkin J.E., Lewis H., Lin S., Neugebauer G.,
Sirota M., Smith G., Ziomkowski C., 1994, \apj 420 L13

\ref
Mazzei P., 1988, Ph.D. thesis, International School for Advanced Studies,
Trieste

\ref
Mazzei P.,  Curir A., Bonoli C., 1995, AJ in press

\ref
Mazzei P., De Zotti G., 1994a, \mnras 266 L5

\ref
Mazzei P., De Zotti G., 1994b, \apj 426 97

\ref
Mazzei P., De Zotti G., Xu C., 1994, \apj 422 81

\ref
Mazzei P., Xu C., De Zotti G., 1992, \apj 256 45

\ref
McCarthy P.J., 1993, \araa 31 369

\ref
Parkes I.M., Collins C.A., Joseph R.D., 1994, \mnras 266 983

\ref
Puget J.L., L\'eger A., 1989, \araa 27 161

\ref
Puget J.L., L\'eger A., Boulanger F., 1985, \aa 142 L9

\ref
Rawlings S., Saunders R., Miller P., Jones M.E., Eales S.A., 1990, \mnras
246 21P

\ref
Rieke G.H., Lebofsky M.J., 1985, \apj 288 618

\ref
Salpeter E.E., 1955, \apj 121 161

\ref
Schmidt M., 1959, \apj 129 243

\ref
Seaton M.J., 1979, \mnras 187 73P

\ref
Serjeant S., Lacy M., Rawlings S., King L.J., Clements D.L., 1995, MNRAS,
in press

\ref
Songaila A., Cowie L.L., Hu E.M., Gardner J.P., 1994, \apjs 94 461

\ref
Spinrad H., Dey A., Graham J.R., 1995, \apj 438 L51

\ref
Trentham N., 1995, MNRAS, in press

\ref
van Ojik R., Roettgering, H.J.A., Miley, G.K., Bremer, M.N., Macchetto F.,
Chambers K.C., 1994, A{\&}A in press.

\ref
Windhorst R.A., Burstein D., Mathis D.F., Neuschaefer L.W., Bertola F.,
Buson L.M., Koo D.A., Matthews K., Barthel P.D., Chambers K.C., 1991,
\apj 380 362

\ref
Windhorst R.A., Gordon J.M., Pascarelle S.M., Schmidtke P.C., Keel W.C.,
Burkey J.M., Dunlop J.S., 1994, \apj 435 577

\ref
Windhorst R.A., Mathis D.F., Keel W.C., 1992, \apj 400 L1

\vfill\eject

\centerline{\bf Figure captions}

\bigskip\noindent
{\bf Fig. 1.} Evolution of the effective optical depth of dust (defined
in \S~2),
of the gas metallicity and of the gas fraction for various choices
of the parameters controlling the SFR and of the lower mass limit of
Salpeter (1955) IMF.

\bigskip\noindent
{\bf Fig. 2.} Distribution of $(B-K)$ and  $(I-K)$ colours
as a function of redshift for galaxies in
the complete ($K<20$) sample of Songaila et al. (1994); stars are
for galaxies lacking redshift measurements, arbitrarily set at $z=5$.
The lines show the relations implied by the present models
for different choices of the SFR, of the redshift of galaxy formation $z_f$,
and of $q_0$. In panels $a$ and $d$, $q_0=0.5$ and $z_f=10$ (except for the
model represented by the dot-dashed line, which has $z_f=2.5$); in panels $b$
and $e$, $q_0=0.5$ and $z_f=5$ (but the dot-dashed line has $z_f=1$);
panels $c$ and $f$ are for $q_0=0$ and $z_f=10$ (but the dot-dashed line has
$z_f=2.5$). In each panel, the solid line refers to $\psi_0 =100\Msyr$
and $n=0.5$, i.e. to dusty early-type models;
the dashed line, to $\psi_0 =50\Msyr$ and $n=0.5$; the dotted line
to $\psi_0 =100\Msyr$ and $n=1$; the dot-dashed line to $\psi_0 =3.5\Msyr$ and
$n=1$, appropriate for late type galaxies.

\bigskip\noindent
{\bf Fig. 3.} Predicted vs observed $K$--$z$ diagram of radiogalaxies
for different values of the galaxy formation redshift, $z_f$, and of $q_0$.
In each panel, the solid line refers to $\psi_0 =100\Msyr$
and $n=0.5$ (dusty early-type galaxies);
the dashed line, to $\psi_0 =50\Msyr$ and $n=0.5$; the dotted line
to $\psi_0 =100\Msyr$ and $n=1$.
The data points are those collected by Eales et al. (1993a;
their Fig. 3). The open symbols correspond to line-corrected magnitudes.

\bigskip\noindent
{\bf Fig. 4.} Spectral energy distribution (SED) of $4C41.17$. All models have
$n=0.5$ [cf. eq. (1)]. The solid line shows the adopted non-thermal SED (see
text). In panel $a$,
model A has age $T= 0.1\,$Gyr, dust temperature $T_d = 46\,$K,
total baryonic mass $m_b = 6.5\times 10^{12}\Msol$,
bolometric luminosity $L_{\rm bol} = 1.2\times 10^{13}\Lsol$, $E(R-K)=1.03$,
gas fraction $f_g=0.90$,  gas metallicity $Z_g = 1.5\times 10^{-3}$,
non-thermal fraction at 1000{\AA} (rest-frame)
$f_{\rm NT} = 0.45$;  model B has $T= 0.1\,$Gyr, $T_d = 46\,$K, $m_b =
8.4\times 10^{12}\Msol$, $L_{\rm bol} = 3.2\times 10^{13}\Lsol$, $E(R-K)=1.3$,
$f_g=0.91$, $Z_g = 3.6\times 10^{-3}$,
$f_{\rm NT} = 0.42$;  model C has $T= 0.05\,$Gyr, $T_d = 60\,$K, $m_b =
1.9\times 10^{12}\Msol$, $L_{\rm bol} = 3.6\times 10^{13}\Lsol$, $E(R-K)=1.63$,
$f_g=0.88$, $Z_g = 0.01$, $f_{\rm NT} = 0.35$.
All models in panel $b$ have $T=2\,$Gyr and $f_{\rm NT} = 1$. The values
of $\psi_0$ and $m_l$ are the same as in panel $a$. Model A has $T_d = 60\,$K,
$m_b = 8.6\times 10^{12}\Msol$, $L_{\rm bol} = 2.8\times 10^{13}\Lsol$,
$E(R-K)=0.90$, $f_g=0.06$; model B has $T_d = 60\,$K,
$m_b = 6.4\times 10^{12}\Msol$,
$L_{\rm bol} = 2.1\times 10^{13}\Lsol$, $E(R-K)=0.90$, $f_g=0.06$; model C has
$T_d = 46\,$K,
$m_b = 4\times 10^{12}\Msol$, $L_{\rm bol} = 2.45\times 10^{13}\Lsol$,
$E(R-K)=0.44$, $f_g=0.0015$. The data points are described in the text.

\bigskip\noindent
{\bf Fig. 5.} SED of B2~$0902+343$. All models have
$n=0.5$. In panel $a$ is $T_d = 46\,$K for all models.  Model A
has $T= 0.3\,$Gyr, $m_b = 5.3\times 10^{12}\Msol$,
$L_{\rm bol} = 2.3\times 10^{13}\Lsol$, $E(R-K)=1.30$,
$f_g=0.77$, $Z_g = 1.3\times 10^{-2}$, $f_{\rm NT} = 0.05$;
model B
has $T= 0.1\,$Gyr, $m_b = 2.5\times 10^{12}\Msol$,
$L_{\rm bol} = 0.97\times 10^{13}\Lsol$, $E(R-K)=1.32$,
$f_g=0.91$, $Z_g = 4\times 10^{-3}$, $f_{\rm NT} = 0.1$;
model C
has $T= 0.05\,$Gyr, $m_b = 1.5\times 10^{12}\Msol$,
$L_{\rm bol} = 2.9\times 10^{13}\Lsol$, $E(R-K)=1.07$,
$f_g=0.88$, $Z_g = 0.01$, $f_{\rm NT} = 0.1$.

In panel $b$, $f_{\rm NT} = 0.6$, $\psi_0 = 100\Msyr$, and $m_l=0.1$,
$T_d = 60\,$K for all models. Model A
has $T= 0.6\,$Gyr, $m_b = 7.9\times 10^{12}\Msol$,
$L_{\rm bol} = 3.5\times 10^{13}\Lsol$, $E(R-K)=1.4$,
$f_g=0.76$, $Z_g = 0.028$; model B
has $T= 0.6\,$Gyr, $m_b = 6.4\times 10^{12}\Msol$,
$L_{\rm bol} = 2.9\times 10^{13}\Lsol$, $E(R-K)=1.3$,
$f_g=0.76$, $Z_g = 0.028$.
The data points are described in the text.

\bigskip\noindent
{\bf Fig. 6.} SED of 6C$1232+39$.
Models A and B have $n=0.5$, model C has $n=1$.
In panel $a$ is $T_d = 60\,K$ for all models.
Model A has $T= 0.1\,$Gyr, $m_b = 3\times 10^{12}\Msol$, $L_{\rm bol} = 1.3
\times 10^{13}\Lsol$, $E(R-K)=1.5$, $f_g=0.91$, $Z_g = 4\times 10^{-3}$,
$f_{\rm NT} = 0.13$;  model B has $T= 0.05\,$Gyr,  $m_b = 2\times
10^{12}\Msol$, $L_{\rm bol} = 4.2\times 10^{13}\Lsol$, $E(R-K)=1.7$,
$f_g=0.88$, $Z_g = 0.01$, $f_{\rm NT} = 0.1$;  model C has $T= 0.2\,$Gyr, $m_b
= 2\times 10^{11}\Msol$, $L_{\rm bol}= 2.6\times 10^{12}\Lsol$, $E(R-K)=0.24$,
$f_g=0.20$, $Z_g = 0.08$, $f_{\rm NT} = 0.09$.
In panel $b$, models A and B have $f_{\rm NT} = 0.6$ and $T_d = 60\,$K. Model A
has $T= 0.8\,$Gyr, $m_b = 4\times 10^{12}\Msol$, $L_{\rm bol} = 2
\times 10^{14}\Lsol$, $E(R-K)=1.35$, $f_g=0.09$, $Z_g = 0.18$;
model B has $T= 0.4\,$Gyr, $m_b =
4\times 10^{11}\Msol$, $L_{\rm bol} = 3\times 10^{12}\Lsol$, $E(R-K)=0.2$,
$f_g=0.05$, $Z_g = 0.15$; model C has
$T= 0.6\,$Gyr, $m_b =
2.2\times 10^{12}\Msol$, $L_{\rm bol} = 1.0\times 10^{13}\Lsol$, $E(R-K)=0.75$,
$f_g=0.57$, $Z_g = 0.03$. The data points are described in the text.

\bigskip\noindent
{\bf Fig. 7.} SED of $53W002$. All models have
$n=0.5$ and $\psi_0=100$. The solid line shows the adopted non-thermal
contribution. Model A has $T= 1\,$Gyr, $m_b = 1.9\times 10^{13}\Msol$,
$L_{\rm bol} = 3.4\times 10^{13}\Lsol$, $E(R-K)=1.35$,
$f_g=0.30$, $Z_g = 0.024$; model B
has $T= 0.6\,$Gyr, $m_b = 1.2\times 10^{12}\Msol$, $L_{\rm bol} =
5.4\times 10^{12}\Lsol$, $E(R-K)=1.5$, $f_g=0.57$, $Z_g = 0.028$;
model C has $T= 1\,$Gyr, $m_b =
1.2\times 10^{12}\Msol$, $L_{\rm bol} = 5.3\times 10^{12}\Lsol$, $E(R-K)=0.84$,
$f_g=0.36$, $Z_g = 0.051$. The data points are described in the text.

\bye